\documentclass[twocolumn,showpacs,preprintnumbers,amsmath,amssymb,superscriptaddress,prl]{revtex4}

\usepackage{dcolumn}
\usepackage{bm}
\usepackage{graphics}
\usepackage{graphicx}
\begin{document}

\title{On The Geometrical Description of Dynamical Stability II}

\author{Eduardo Cuervo-Reyes}
\thanks{E. Cuervo-Reyes and Ramis Movassagh contributed equally to this work}
\affiliation{LAC ETH-Zurich, Switzerland}

\author{Ramis Movassagh}
\email[Corresponding author: ]{movassagh@collegium.ethz.ch }
\affiliation{LAC ETH-Zurich, Switzerland}
\affiliation{Collegium Helveticum, Zurich, Switzerland }

\date{\today}
\begin{abstract}
Geometrization of dynamics using (non)-affine parametization of arc length with time is investigated. The two archetypes of such parametrizations, the Eisenhart and the Jacobi metrics, are applied to a system of linear harmonic oscillators. Application of the Jacobi metric results in positive values of geometrical lyapunov exponent. The non-physical instabilities are shown to be due to a non-affine parametrization. In addition the degree of instability is a monotonically increasing function of the fluctuations in the kinetic energy. We argue that the Jacobi metric gives equivalent results as Eisenhart metric for ergodic systems at equilibrium, where number of degrees of freedom $N\rightarrow\infty$. We conclude that, in addition to being computationally more expensive, geometrization using the Jacobi metric is meaningful only when the kinetic energy of the system is a positive constant.

\end{abstract}


\maketitle

Geometrization of dynamics is an alternative for stability analysis of Hamiltonian systems, when normal lyapunov exponent is not\cite{Eisenhart,Rand}. Its non-perturbative nature has opened the possibility for studying general conditions under which the traditional tools \cite{Poincare}, suitable for quasi-integrable systems, cannot be applied due to the strong mixing terms among the different degrees of freedom. Moreover, it seems to be a promising framework for setting the statistical mechanics on dynamical and geometrical grounds. The general idea is to map the real motions $q^i(t)$ of the system \begin{eqnarray}
H(\bm{q},\bm{p})&=&\frac{1}{2}a^{ij}(\bm{q})p_{i}p_{j}+V(\bm{q}) \\
p_{i}&=&a_{ik}(\bm{q})\dot{q}^k \nonumber
\end{eqnarray}
\noindent as geodesics on a properly defined manifold. Then, the quantitative stability analysis is obtained by means of the  evolution with respect to arc length of the vector field of the geodesic spread $\xi^i(s)\equiv \left[\frac{\partial\gamma^i(\tau,s)}{\partial\tau}\right]_{\tau=\tau_0}$. $\gamma^i(\tau,s)$ is a congruence of geodesics, $s$ is the arc length and $\tau$ parametrizes the different geodesics.
The general expression for $\xi$ dynamics is given by the Jacobi-Levi-Civita (JLC) equation\cite{Levi}
\begin{equation}
\frac{D^2\xi^i}{ds^2}+R^i_{jlk}\frac{dq^j}{ds}\xi^l\frac{dq^k}{ds}=0\label{JLC}
\end{equation}
\noindent Where $\frac{D}{ds}$ and $R^i_{jkm}$ are respectively the covariant derivative along the geodesic and the Riemann-Christoffel tensor. Equation (\ref{JLC}), after opening the covariant derivatives, and taking into account the general expression for the geodesic equation 
 \begin{equation}
\frac{d^2q^k}{ds^2}+\Gamma^k_{lj}\frac{dq^l}{ds}\frac{dq^j}{ds}=0
\end{equation}
\noindent becomes 
\begin{equation}
\frac{d^2\xi^k}{ds^2}+2\Gamma^k_{lj}\frac{dq^l}{ds}\frac{d\xi^j}{ds}+\Gamma^k_{lm,j}\frac{dq^l}{ds}\frac{dq^m}{ds}\xi^j=0\label{JLCopen}
\end{equation}

From the integration of Eq.\ref{JLCopen}, the geometrical indicator of stability is obtained:
\begin{equation}
\lambda=\lim_{s\rightarrow \infty}\frac{1}{s}\log\left(\frac{||\xi(s)||}{||\xi(0)||}\right)
\end{equation}

 Two widely used frameworks entail the application of the Eisenhart and the Jacobi metrics; the latter is given by $(g_J)_{ij}\equiv 2[E-V(\bm{q})]a_{ij}(\bm{q})$ and it is only applicable for time independent systems\cite{Pettini1,Pettini3}.

 Usage of the Jacobi metric in the past has given the surprising suppression of chaos with increasing number of degrees of freedom\cite{Pettini2}. This was linked to the approximations done in the formalism; furthermore, parametric resonance was considered to be the fundamental source of instability arising from the fluctuating curvature. In a previous work \cite{ramed1} we addressed these controversial results aiming to warn about which conditions must be fulfilled in order to expect consistent results using either of the two metrics. In the context of a two dimensional system, we argued that the non-affine parametrization of arc length ($s$) with time ($t$) can be a significant source of non-physical instability. The danger of non-affine parameterization seems to reside in the fact that trajectories are not compared at the same time and it is manifested geometrically by the large amplitude oscillations of the curvature.  Moreover, from a mathematical point of view, it has been shown that due to the degeneracy of the Jacobi metric in the boundary set, there are geodesics that do not correspond to physical motions\cite{marek2}.  

In this letter we aim to make the consequences of the non-affine parametrization transparent and precise. Therefore, we use a well known paradigm of stability: a system of independent harmonic oscillators. We apply the stability analysis using Jacobi metric; any evidence of chaos in this dynamical system can then be understood as a failure of the method. Specifically, we investigate the dependence of the results obtained for the indicator of (in)stability on the fluctuation of the kinetic energy along the geodesics. In contrast we will show analytically that the stability results in the framework of Eisenhart metric are in accord with physical expectations.

\begin{figure}[h]
%\scalebox{.55}{\includegraphics{polar_lyapun.eps}}
\scalebox{.55}{\includegraphics{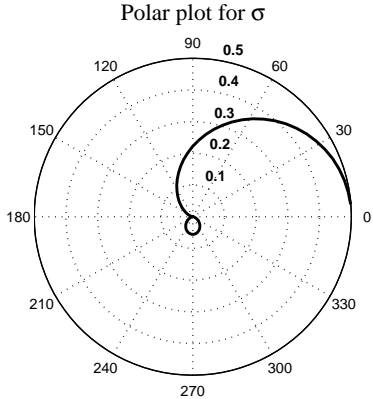}}
\caption{\label{fig1} The polar plots for the square of the fluctuations in the kinetic energy $\sigma$ vs. the fraction of the phase circle $f$ for $N=10$, $\omega = 2\pi$.}
\end{figure}

The time-independent Hamiltonian 
\begin{eqnarray}
H(\bm{q},\bm{p})&=&\frac{1}{2}\left(\delta^{ij}p_{i}p_{j}+ \omega^2\delta_{ij}q^iq^j\right) \label{Hamilton}\\
p_{i}&=&\delta_{ik}\dot{q}^k\nonumber
\end{eqnarray}
is our basic model. The solutions to the equations of motion are
\begin{equation}
q^k(t)=C^k\cos(\omega t+\theta_k) \label{dysol}
\end{equation}
\noindent where $C^k$ and $\theta_k$ depend on the initial conditions and in our case were chosen as $C^k=1$ and $\theta_k=k\frac{2\pi f}{N}$, $k=1,...,N$. Here $N$ stands for the number of degrees of freedom and the phases, $\theta_k$, are homogeneously distributed on a fraction ($f$) of $2\pi$ (phase circle). Using Eq. (\ref{dysol}) we get for the kinetic energy $T$ and the square of its fluctuation
\begin{eqnarray}
T&=&N\left(\frac{\omega C}{2}\right)^2\left[1-\sqrt{2\sigma}\cos(2\omega t+2\pi f\frac{N+1}{N})\right] \\
\sigma&\equiv&\frac{<T^2>-<T>^2}{<T>^2}=\left(\frac{\sin(2\pi f)}{\sqrt{2}N\sin(2\pi f/N)}\right)^2\label{fluct}
\end{eqnarray}
These settings allow us to tune the fluctuation $\sqrt{\sigma}$, varying the values of $f$ and $N$. 
It can be easily seen from the above equations that $\sigma$ decreases  with increasing $N$, having  the limit $|\frac{\sin(2\pi f)}{\sqrt{2}2\pi f}|$ as $N\rightarrow\infty$. When all of the oscillators are in phase ($f=0$), the variance takes its maximum value and the kinetic energy reaches the zero value every $\Delta t=\frac{\pi}{\omega}$.
The specific dependence of $\sigma$ on the fraction of the phase circle for $N=10$ is shown in FIG.\ref{fig1}. This is of interest under statistical considerations, where one expects the fluctuations of the kinetic energy to decrease as one makes more use of the phase circle.
 Generally in an $N$-body bounded system, the kinetic energy is a fluctuating quantity. In non-integrable systems due to the mixing terms (non-separability), these fluctuations are normally reduced with increasing number of degrees of freedom. Since our toy model has no interactions at all and is coherent, it is additive but not necessarily ergodic; therefore $\sigma$ does not vanish when $N\rightarrow\infty$. However, our main interest is to explicitly show the shortcomings of the Jacobi metric, which is unaffected by the properties of our toy model. Moreover, we will show that $\lambda$ strictly depends on $\sigma$.

Eq.\ref{JLC}, when applied to Eq.\ref{Hamilton} using Jacobi metric, becomes
\begin{widetext} 
\begin{equation}
\frac{d^2\xi^k}{dt^2}+\omega^2\xi^k + \frac{\omega^2}{T}\delta_{lj}\left[\left(q^k{\dot q}^l-q^l{\dot q}^k\right)\frac{d\xi^j}{dt}+\left(\omega^2q^kq^l-{\dot q}^k{\dot q}^l-\frac{\omega^2}{T}\delta_{im}{\dot q}^iq^m{\dot q}^kq^l\right)\xi^j\right]=0 \label{xidyn}
\end{equation}
\end{widetext}
This is radically different (due to the extra mixing terms) to that obtained from Eisenhart's metric, which corresponds to the tangent dynamic equations of motion
\begin{equation}
\frac{d^2\xi^k}{dt^2}+\omega^2\xi^k=0\label{eisen}
\end{equation}
It is perhaps worth while mentioning that the solution of Eq.(\ref{eisen}) evidently gives $\lambda \leq 0 $ for any setting of initial conditions; moreover $\lambda=0$ is of measure zero. Thus, the Eisenhart geometrization behaves properly for this system and is a good indicator of (in)stability.
Once the analytic solutions (\ref{dysol}) were substituted in Eq.(\ref{xidyn}) it takes the final form
\begin{equation}
\frac{d^2\xi^k}{dt^2}+\omega^2\xi^k + \omega I^k_j\frac{d\xi^j}{dt}+\omega^2\left[J^k_j+K^k_j\right]\xi^j=0 \label{fin}
\end{equation}
\noindent with the couplings $I^k_j$, $J^k_j$ and $K^k_j$ given by    
%\begin{widetext}
\begin{subequations}
\begin{eqnarray}
 I^k_j&=&\frac{\omega^2 C^2}{T}\sin(\theta_k-\theta_j)\\
 J^k_j&=&\frac{\omega^2 C^2}{T} \cos(2\omega t+\theta_k+\theta_j)\\
 K^k_j&=&-\frac{\omega^4 C^4}{2T^2}\sin(2\omega t +2\pi f\frac{N+1}{N})\times\\
&&\left[\sin(2\omega t+\theta_k+\theta_j)-\sin(\theta_k-\theta_j)\right] \nonumber
\end{eqnarray} 
\end{subequations}
%\end{widetext}
The above expressions have a typical time dependence and structure from which instabilities such as parametric resonance may arise \cite{Landau}, with the extra ingredient that the amplitude of the couplings are generally unbounded due to the inverse powers of $T$. Although these expressions are system dependent, the appearance of inverse powers of $T$ is a general feature of the Jacobi metric and it is responsible for the singularities in the boundary set.

\begin{figure}[h]
\scalebox{.5}{\includegraphics{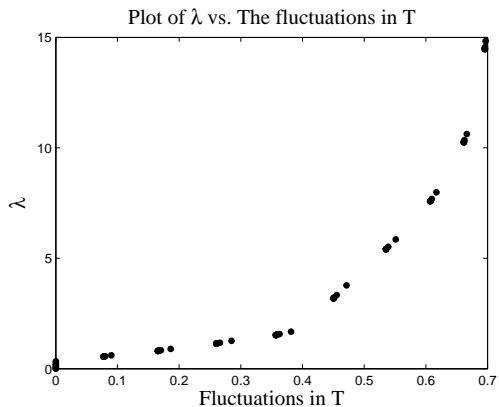}}
\caption{\label{fig2} The superposition of the geometrical lyapunov exponent $\lambda$ vs. the fluctuations of the kinetic energy $\sqrt{\sigma}$ for $ N=2+j^2$ and $f=0.05,0.1,...,0.45$, where $j=1,...,14$.} 
\end{figure}

In order to make our analysis quantitative, $\lambda$ was obtained from the numerical integration of $\xi$ dynamics Eq.(\ref{fin}) computed using different values of $f$ and $N$. In FIG.\ref{fig2} we superimpose these values of $\lambda$ as functions of $\sqrt{\sigma}$ (i.e. fluctuations in the kinetic energy), obtained from the Eq.\ref{fluct} for different settings. The fact that the plot shows a smooth dependence implies that the relevant quantity of interest, on which $\lambda$ depends, is the fluctuation of the kinetic energy. 

\begin{figure}[h]
\scalebox{.4}{\includegraphics{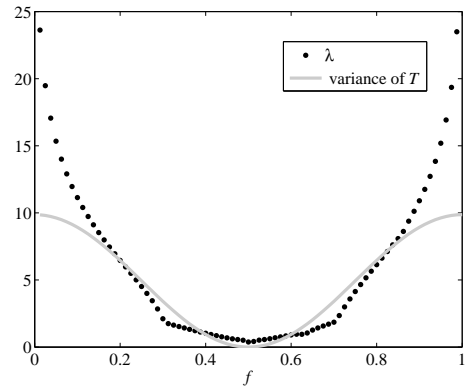}}
\caption{\label{fig3} Comparison of $\lambda$ and the absolute variance for two degrees of freedom vs. $f$. }
\end{figure}

\begin{figure}[h]
\scalebox{.5}{\includegraphics{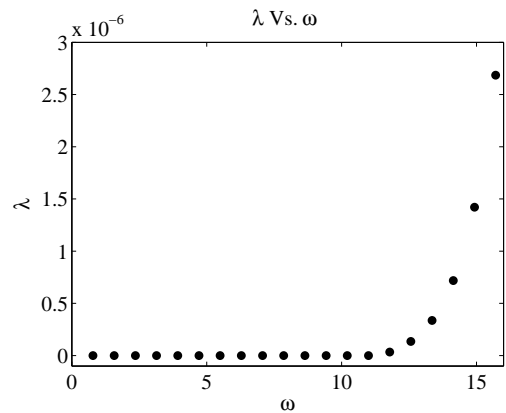}}
\caption{\label{fig4} The geometrical lyapunov exponent $\lambda$ vs. the fundamental frequency of oscillation $\omega$ for $N =10$ and $f=1$.}
\end{figure}

In the light of a previous work\cite{ramed1}, FIG.\ref{fig3} makes evident the strong correlation between the values of the geometrical stability indicator $\lambda$ and the variance in the kinetic energy with initial conditions for the two oscillators. Needless to say, strong non-physical divergences arise from the ``kicks'' produced in the $\xi$-dynamics when the kinetic energy approaches zero periodically. As we discussed before, these ``kicks'' are nothing but divergences of the curvature tensor close to the boundary set because the curvature tensor contains inverse powers of the kinetic energy. 
Lastly we show, for the maximum uniform distribution, i.e.$f=1$, the dependence of $\lambda$ on fundamental frequency of the oscillators $\omega$. As is evident in the FIG.\ref{fig4}, the lyapunov exponent increases with increasing $\omega$. Curiously, although $\sigma=0$ according to Eq.\ref{fluct}, we obtained small but positive $\lambda$. The reason is that equations like (\ref{fin}) have extremely unstable behavior. Therefore, even in cases where the variance was set to zero, small fluctuations produced by numerical evaluation are amplified, leading to positive exponents, that are naturally reduced with reducing the value of ($\omega$), which multiplies the mixing terms in Eq.\ref{fin}.
 
In this letter we have shown that in the framework of the Jacobi metric the evolution of the vector field of the geodesic spread is not physically meaningful. $\xi$-dynamics shows an exponential divergence $\lambda>0$, for a system which is physically stable.  We showed that  these non-physical divergences are due to the fluctuations in the kinetic energy, which for ergodic systems should vanish at the thermodynamical limit. In addition these oscillations have the cumulative effect of parametric resonance-type that can be confused with real instability. This instability was induced by the non-affine parametrization of the arc length with time which causes the local comparison of the trajectories to be at different times.  It is expected that, only for ergodic systems at the limits of $N\rightarrow\infty$ and $t\rightarrow\infty$, the Jacobi metric would give equivalent results as the Eisenhart metric would. Moreover, the stability of this system of independent oscillators is shown to be dependent on the fundamental frequency, which of course is incorrect. Instabilities show up even in cases where the variance of $T$ is theoretically zero by the choice of the initial conditions. It is an evidence of the unstable character of the solutions of the JLC equation for the Jacobi metric, which amplify the error introduced by numerical integration.  Therefore, the Jacobi metric gives correct stability results only when the kinetic energy of the system as whole is strictly a positive constant, which is practically impossible.  

We thank Reinhard Nesper for discussions and his support. This project was funded by Swiss National Science Foundation and The Cogito Foundation.

\end{document}